# A Comprehensive Model of Nitrogen-Free Ordered Carbon Quantum Dots

D.W. Boukhvalov[1,2,*], V.Yu. Osipov[3], B.T. Hogan[4,5], A. Baldycheva[4]

[1]College of Science, Institute of Materials Physics and Chemistry, Nanjing Forestry University, Nanjing 210037, P. R. China
[2]Institute of Physics and Technology, Ural Federal University, Mira 19 str., 620002, Yekaterinburg, Russia
[3]Ioffe Institute, Polytechnicheskaya 26, St. Petersburg 194021, Russia
[4]STEMM Laboratory, University of Exeter, Exeter, UK
[5]Department of Electrical and Information Engineering, University of Oulu, Oulu, Finland

*We propose and demonstrate a novel range of models to accurately determine the optical properties of nitrogen-free carbon quantum dots (CQDs) with ordered graphene layered structures. We confirm the results of our models against the full range of experimental results for CQDs available from an extensive review of the literature. The models can be equally applied to CQDs with varied sizes and with different oxygen content in the basal planes of the constituent graphenic sheets. We demonstrate that the experimentally observed blue fluorescent emission of nitrogen-free CQDs can be associated with either small oxidised areas on the periphery of the graphenic sheets, or with sub-nanometre non-functionalised islands of $sp^2$-hybridised carbon with high symmetry confined in the centres of oxidised graphene sheets. Larger and/or less symmetric non-functionalised regions in the centre of functionalised graphene sheet are found to be sources of green and even red fluorescent emission from nitrogen-free CQDs. We also demonstrate an approach to simplify the modelling of the discussed $sp^2$-islands by substitution with equivalent strained polycyclic aromatic hydrocarbons. Additionally, we show that the bandgaps (and photoluminescence) of CQDs are not dependent on either out-of-plane corrugation of the graphene sheet or the spacing between $sp^2$-islands. Advantageously, our proposed models show that there is no need to involve light-emitting polycyclic aromatic molecules (nanographenes) with arbi-*

*trary structures grafted to the particle periphery to explain the plethora of optical phenomena observed for CQDs across the full range of experimental works.*

E-mail: danil@njfu.edu.cn

## 1. Introduction

Carbon dots are an exciting member of the family of nanocarbon materials, diverging in properties from the well-known nanographene [1]. They have typical sizes of less than 8-10 nm, but their compact nanoscale skeletons can nevertheless be structurally diverse. Carbon dots were first investigated by Xu *et al.* in 2004 [2]. Subsequently, the first bright luminescent carbon nanoparticles, called quantum-sized carbon dots or carbon quantum dots (CQDs), were obtained in 2006 by Sun *et al.* [3]. Due to their plethora of distinctive properties— including an inexpensive synthesis route, bright photoluminescence, excellent solubility in water, and potential for various useful features of the chemical nature of the surface— carbon dots have attracted increasing global research recognition with proposed applications as wide-ranging as biomedical techniques (as fluorescent markers) or alternative light sources [4-11]. However, the abundance of possible experimental materials associated with the variety of structural forms of CQDs, and the insufficient development of current theoretical models (reviewed herein) are acting as a significant brake on further meaningful work in these directions. The diversity of the inorganic, organic, and biomolecular precursors used for manufacturing CQDs (from plant seeds and biowaste, to graphite and carbon fibres, see Table I [12,13]), the various sizes and morphologies enabled, and the presence of nitrogen (as either a dopant or a contaminant) or other substitutional heteroatoms are all barriers to the generalisation and systematisation of our knowledge on CQDs. Further development towards practical applications of CQDs would benefit profoundly from systematic knowledge of the relations between physical properties and methods of production, atomic structures, and morphological properties of the CQDs. It is notable that only a small fraction of the literature using the term CQDs reports nano-objects with luminescence spectra typical of QDs. Rather, the majority of existing work reports carbon nano-objects with optical properties similar to bulk structures. Therefore, herein we use the term CQDs to represent all nanosized carbon particle-like structures with distinct luminescent properties.

*Structure.* CQDs can be classified based on their specific properties, structures, and the functional groups attached to their surfaces [14]. CQDs can have structures analogous to core-shell nanoparticles. In this case, the 'core' contains carbon atoms in both $sp^2$ and $sp^3$ hybridisations (similar to amorphous carbon). Various functional groups on the carbon surface (predominantly oxygen-containing aldehydes, hydroxyls, epoxies, carbonyls, and carboxyls) play the role of the 'shell'. These groups come either from precursor organic compounds such as citric acid (widely used for bottom-up fabrication of CQDs [16]) or appear due to oxidation during the unzipping and top-down cutting of larger nanoscale carbon bodies (nanotubes, graphite, *etc.*). [17] Additionally, it is assumed (although widely explored, this has not yet been confirmed by the data available in the literature) that carbon dots may contain other molecular fragments on the surface (*e.g.*, planar molecules or embedded $sp^2$-hybridised carbon domains consisting of up to 18 conjugated aromatic rings) which can fluoresce in the range from 400 to 700 nm [18].

In general, the family of CQDs includes at least three classes of nanoparticles: graphene ordered carbon dots (denoted o-CD herein), carbon nanodots (CNDs), and polymer dots (PDs) [19]. Particles of all these types have approximately the same photophysical and photoelectrochemical properties but differ in their internal structures and the composition of functional groups on their surfaces. The core-shell structure is a prominent feature for CNDs and PDs, but is less relevant for o-CDs consisting of a few planar graphenic layers with round shapes and a well-defined stacking order.

PDs typically have larger sizes (up to 20-30 nm) and a C-H-O-N type atomic composition, where the first three elements occur approximately equally, and the nitrogen content can vary [20,21]. They are usually assembled from linear polymer or monomer species [19]. The key feature of PDs is the comparatively large content of hydrogen and oxygen atoms.

Unlike the other types of CQDs, CNDs are considered to be quasi-spherical carbonic nanoparticles which mainly consist of an amorphous $sp^3/sp^2$ structural core [22]. Such particles (even when less than 10 nm in size) can also contain embedded $sp^2$-hybridised nanocrystalline species [14]. It is difficult to observe lattice fringes in CNDs due to their overall amorphous structure and the non-crystalline structure of their elementary subunits [23, 24].

o-CD particles contain a single or multilayer graphite-like structure across which various other functional groups can be distributed. The presence of large numbers of functional groups at the edges or localised at defects in the interlayer regions (for example, on vacancies and multivacancy clusters) endows the non-functionalised parts of the graphenic layers of o-CD particles with additional quantum restrictions and edge effects. High resolution TEM images of such particles have demonstrated that they are highly crystalline and well-ordered with interplanar distances of around 0.21 nm, corresponding to the (110) lattice spacing of graphene, and spacing between graphenic layers of around 0.34 nm or slightly more [23]. The $d_{002}$ spacing (0.403 nm) can be broadened up to 0.41 nm through the introduction of oxygen-related groups into the interlayer space of nanoparticles *via* chemical oxidation [23, 24]. Typically, o-CD particles consist of at least four or five stacked graphenic layers (nanodiscs).

*Modelling and prediction of properties*. The standard approach to the modelling of CQDs is to propose a more or less realistic atomic structure, and then to iterate structural changes until coincidence of the predicted optical properties with those experimentally observed is obtained. There are two possible approaches to building useful models of nanoparticles. The first is to consider the local structure of the nanoparticles' surfaces as a slab or monolayer within periodic boundary conditions. This approach works rather well for nanoparticles larger than 10 nm, since contributions from the edges are much smaller than those from bulk-like areas (see, *e.g.*, [25]). The second approach is the construction of finite element nanoclusters. In the case of CQDs, the model system is built of single or stacked polycyclic aromatic hydrocarbons (such as coronenes, also called nanographenes) or diamond-like clusters (see, *e.g.*, [26]). The latter approach has two potential advantages: (i) the possibility of directly taking into account contributions from the edges, and (ii) the possibility of exploiting modern quantum chemistry-based programmes which can be used to model molecule-like systems. Using this approach, time-dependent density functional theory (TD-DFT) based tools can be used to describe the optical properties of the systems. This approach provides a good description for CQDs constructed from a carbon core with *$sp^3/sp^2$* structure and light-emitting surface molecular structures composed of up to five conjugated aromatic rings (see, *e.g.*, [27,28]). Recently the same approach was used to explain unprecedented ultraviolet-B luminescence of carbon dots consisting of ordered graphenic layers due to fluorescent configurations of specially conjugated three benzene rings attached to their surfaces [29]. However, a significant disadvantage of this approach is the arbitrary choice of the sizes of fluorescent molecular structures and especially the shapes of the clusters built from benzene rings. In some cases, clusters of different shapes and sizes can have similar HOMO-LUMO gap values. The same arguments can be applied to the explain the optical properties of CQDs by the presence of functional groups at the edges or by the special shape of the edges of nanographenes or molecules [26-28]. The lack of reliable experimental feedback about the size, shape, and edge structures of the nanocarbons makes it impossible to distinguish robust models from those built on sand. As such, this model of light-emitting molecular structures is of limited use; it works well in specific cases but does not generalise to most situations. This is especially true for carbon dots, consisting of stacked graphene layers (nanodiscs), with bandgaps smaller than the energy of fluorescent radiation.

For the case of CQDs of a few nanometres in size and with graphite-like structures (o-CD), all of the discussed approaches are significantly lacking. Increasing the size of pure nanographenes (the core part) leads to a gradual red shift of the predicted photoluminescence wavelength, with the energy gap eventually vanishing for nanographenes with sizes greater than 1.2 nm. On the other hand, multiple experimental results demonstrate the presence of photoluminescence (PL) in the blue region of the spectrum (400~450 nm) in o-CDs with sizes around 1-2 nm (see Tab. I). The blue shift of the PL should correspond with a decrease in the size of the related nanographenes, but the experimental works summarised in Tab. I do not show the presence of any sub–nanometre nanographenes in the studied systems.

The greatest limitation of the existing modelling approaches described is the absence of predictive power. These approaches cannot instruct one on how to change the production method or treatment of CQDs to obtain desired properties for the modelled system. With this in mind, the development of a more robust approach to the modelling of o-CDs is fundamental to the systematisation of experimental results and to guide further synthesis of o-CDs towards specific applications in a controlled manner.

The purpose of this work is to elucidate a structural model for o-CD particles. Initially, we consider nitrogen–free o-CDs with graphene-like morphology (nanographenes, larger graphene sheets, graphenic stacks, *etc.*); specifically, those that exhibit regular fringes in their TEM images and emit light in the range from 400 to 700 nm. This is because it is currently unclear why relatively large o-CD particles up to 5–6 nm in size emit blue light; in TEM images they appear as a stack of regularly arranged graphene layers whereas such a structure would not be expected to emit high-energy blue light unless there were carbon dot particles with an amorphous mixed *$sp^2$/$sp^3$* structure.

## 2. Review of Experimental and Theoretical Studies of Nitrogen-free CQDs

Although a considerable number of experiments have been reported nitrogen-free CQDs, the number of works which comprehensively describe the structure and properties of their investigated systems is rather small (see Table I). We consider five key elements of the reported experimental results: (i) the general approach to their production; (ii) the ratio of $sp^2$ to $sp^3$ hybridisation of the carbon content; (iii) the mean size of the CQDs; (iv) whether the CQDs have a layered structure; and (v) the wavelength of the photoluminescence peak [30-46].

The production approach can be categorised as either top-down or bottom-up. In the top-down approach, graphite, graphene oxide, or carbon fibres are reduced in size by physical and/or chemical treatments. In the bottom-up approach, graphene sheets are formed by chemical and physical assembling of fragments of organic molecules obtained from other carbon-containing materials. Top-down CQD production results in the formation of highly ordered layered structures with crystalline fringes that can be seen in HRTEM (high resolution transmission electron microscopy) images. With the bottom-up approach, the sizes of fabricated CQDs are usually smaller and HRTEM images often indicate the absence of a layered structure (see compounds 2–4 in Table I).

X-ray Photoelectron spectroscopy (XPS) is a powerful tool for describing the chemical composition of carbon-based systems. Carbon atoms in a graphenic sheet can be either $sp^2$ or $sp^3$ hybridised. $sp^2$ hybridisation corresponds to either: non-functionalised carbon atoms in central part, or carbon atoms at the edges passivated by monovalent groups such as hydrogen or hydroxyl. $sp^3$ hybridisation corresponds to either: functionalised atoms in the central part of graphene, or carbon atoms from carbonyl and carboxyl groups [47]. A simple estimate shows that, for a 3–4 nm nanographene, the ratio of carbon atoms in the central part to those along the edges is approximately 5:1. In the XPS spectra of C 1s, both C=O and –COOH groups have distinct peaks

(at about 288 and 289 eV, respectively) [48] separated by more than 1 eV from the peaks associated with various $sp^2$ and $sp^3$ carbon atoms located in the central part of graphene [49-51]. All the experimental results summarised in Table I show an insignificant number of carboxyl and carbonyl groups and a significant contribution from functionalised carbon atoms with $sp^3$ hybridisation located away from the edges. Thus, the second important descriptor for the CQDs is the ratio of $sp^2$ to $sp^3$ carbon atoms located away from the edges of the graphenic sheets. This ratio can be directly obtained from XPS measurements. The spectra are also useful for evaluating the ratio of oxidised and non-oxidised species on the graphenic sheets, which can then be further used in modelling the system (see, *e.g.*, [52]).

**Table I.** Summary of methods of production, source of carbon, $sp^2/sp^3$ ratio of carbon atoms in central parts of the graphenic sheets as estimated from X-ray Photoelectron Spectroscopy (XPS), mean carbon quantum dot (CQD) size, presence or absence of an observed layered structure in TEM images, and wavelength of the photoluminescence maximum for experimentally characterised nitrogen-free CQDs.

| Method | ID | Source of carbon | $sp^2/sp^3$ ratio | Mean size (nm) | Layered structure? | Maximum PL wavelength (nm) | Colour |
|---|---|---|---|---|---|---|---|
| Bottom-Up | 1 | fennel seeds [30] | 3:1 | 6.0 | Yes | 420 | violet |
| | 2 | fenugreek seeds [31] | 5:1 | 5.0 | No | 410 | violet |
| | 3 | chia seeds [32] | 1:7 | 3.8 | No | 420 | violet |
| | 4 | tapioca powder [33] | 1:5 | 3.5 | No | 500 | cyan/green |
| | 5 | lemon juice [34] | 4:1 | 2.2 | Yes | 482 | blue/cyan |
| | 6 | trisodium citrate [35] | 4:1 | 1.2 | Yes | 420 | violet |
| | 7 | isopropanol [36] | 4:1 | 1.5 | | 400 | violet |
| | 8 | 1,3-Dihydroxynaphthalene [37] | 20:1 | 5.0 | Yes | 600 | red/orange |
| | 9 | graphite [38] | 3:2 | 3.5 | Yes | 600 | red/orange |
| | 10 | CF [39] | 3:1 | 3 | Yes | 400 | violet |
| | 11 | GO [40] | 4:1 | 2—5 | Yes | 450 | violet/blue |
| | 12 | CF [41] | 2:1 | n/a | | 450 | violet/blue |
| Top-Down | 13 | GO + benzene alcohol [42] | 2:1 to 20:1 | 6.2 | Yes | 420 | violet |
| | 14 | GO + organic acids [43] | 20:1 | 11 | Yes | 560 | green/yellow |
| | 15 | GO [44] | 10:1 | 4 | Yes | 410 | violet |
| | 16 | graphite [45] | various | 4 | Yes | Various | — |
| | 17 | graphite, nanotubes [46] | 10:1 | 2.5~3 | Yes | 540, 460 | cyan/blue |

Note that in almost all the layered systems presented in Table I (except 15), the approximate number of carbon atoms in *sp²* hybridisation is three or more times larger than the number of oxidised carbon atoms with *sp³* hybridisation. All previous theoretical and experimental considerations of the relationship between oxidation level and bandgap in GO demonstrated that reduction of GO (*i.e.*, increasing the *sp²*/*sp³* ratio to 3:1) results in the vanishing of the bandgap [53].

Finally, we consider the characteristic wavelength (or photon energy) of the PL emission. The spectral maxima of PL emission are in the range 400–450 nm (corresponding to blue light emission) or ten of the fifteen compounds considered in Table I. Note that the similar values of the PL spectral maxima were recorded for CQDs of different sizes and with different *sp²*/*sp³* ratios. For example, compounds 1, 6, and 13 have spectral maxima at ~420 nm but their sizes vary from 1.2 to 6 nm and *sp²*/*sp³* ratios from 2:1 to 20:1. On the other hand, the CQDs with *sp³* content larger than *sp²* content (2) or similar to it (15) also demonstrate PL within the same or almost the same spectral range. These various similarities and dissimilarities provide us enough information to start the simulation of realistic nitrogen-free graphenic CQDs.

Recent theoretical studies report systematic investigations of nanographenes of various shapes, sizes, and edge structures as possible models for some CQDs [54-57]. The effects of adatoms [58] and defects on the edges [40,43,45,59] were also simulated. An important finding of these studies was a rapid decay in the value of the bandgap as the size of nanographenes increased, with the energy gap closing for nanographenes larger than 1 nm. However, this repeatedly reproduced theoretical result is inconsistent with numerous experimental observations of PL in CQDs with layered structures of a few nanometres in size (see Table I). Another common shortcoming of the cited theoretical studies in terms of their application to CQDs is the neglect of the contribution from functionalised regions in the central parts of the graphenic sheets— a feature observed in all CQDs. Thus, based on a brief observation of these recent theoretical results, we hypothesise that simulation of partially functionalised large graphenic sheets may provide a better description of real o-CDs.

## 3. Methods and Terminology

Theoretical modelling was carried out using SIESTA pseudopotential code [60] employing the generalised gradient approximation (GGA-PBE) [61] for the exchange-correlation potential in a spin-polarised mode. A full optimisation of the atomic positions was carried out during which the electronic ground state was consistently found using norm-conserving pseudopotentials [62] for the cores with a double-$\xi$-plus polarisation basis for the localised orbitals of non-hydrogen atoms and a double-$\xi$ for hydrogen atoms. The forces and total energies were optimised with an accuracy of 0.04 eV Å$^{-1}$ and 1.0 meV/cell (or less than 0.02 meV/atom), respectively.

Herein, we will discuss the relationships between atomic structures of various simulated systems and the value of the energy gap between the valence band maximum (VBM) and con-ductive band minimum (CBM) in periodic systems (also called the highest occupied and lowest unoccupied molecular orbitals, respectively). The value of this bandgap defines the PL wavelength. The transition from computationally expensive state-of-the-art many-body model calculations to DFT corresponds with the description of the motion of each electron as motion in the mean field of all other electrons. The electron mean field acting on a single electron is the electrostatic field created by all electrons in the system, including the considered electron. This simplification allows calculations for realistic (complex) systems but has some disadvantages such as underestimation of the bandgap [63]. This disadvantage of the DFT-based methods can be fixed by using the GW (Green's function, screened Coulomb interaction) approximation approach [64], using so-called hybrid functionals [65], or by estimation of the real bandgap from that calculated within the standard DFT framework. Since at the current level of hardware development, calculation of hundreds of atoms using GW methods is prohibitively computationally costly, and hybrid functionals are not implemented in the SIESTA code, we apply the third approach herein using the relationships between the values of the bandgap calculated using standard DFT and GW approaches reported by van Schilfgaarde *et al.* [64] for a broad range of values (from 0 to 7 eV). We further report and discuss the values of the bandgap calculated using standard DFT, and in some cases discuss the relationships between the calculated and real bandgaps. For further reference, the relationships between PL colours, photon energy (real bandgap), and approximate values of the calculated bandgap are summarised in Table II.

**Table II.** Summary of PL colours, their related measured photon energies (corresponding with the real bandgap), and approximate values of the bandgap calculated using the standard DFT-GGA approach.

| Colour | Photon Energy (eV) | Calculated Bandgap (eV) |
|---|---|---|
| Violet | 2.75-3.26 | 1.9-2.1 |
| Blue | 2.56-2.75 | 1.7-1.9 |
| Cyan | 2.48-2.56 | 1.6 |
| Green | 2.19-2.48 | 1.3-1.5 |
| Yellow | 2.10-2.19 | 1.2 |
| Orange | 1.98-2.10 | 1.1 |
| Red | 1.65-1.98 | 0.6-1.0 |

## 4. Results and Discussion

### 4.1. Building a Proper Model for ordered Carbon Dots

Based on the experimental data summarised in Table I, we can assume that most o-CDs constructed from graphenic layers have lateral sizes greater than two nanometres (predominantly 4–5 nm). Thus, the contribution from edge states is significantly less than the contribution from central areas. In addition, XPS measurements clearly demonstrate the presence of a sufficient amount of carbon in $sp^3$ hybridisation associated with the presence of hydroxy and epoxy groups in the central part of the graphenic sheets, along with a negligible contribution from carbonyl and carboxyl groups on the edges. Infrared absorption measurements also demonstrate the significant presence of hydroxyl groups (see, *e.g.*, Wei *et al.* [45]) which cannot be related only with the periphery of graphene sheets of several nanometres in size. Based on the combination of the large size and small edge state contributions observed in the majority of real *o*-CDs considered in Table I, we choose to simulate o-CDs as supercells of partially functionalised graphene within periodic boundary conditions (Fig. 1a-d). This model provides a correct description of the state of basal planes in layered structures of a few nanometres in size. Note that this model does not account for the contribution of edge groups for the reasons discussed in the introduction. The interlayer interactions are also omitted; in graphitic nanoparticles these interactions are rather weak and many inorganic intercalants (*e.g.,* water molecules) can penetrate between the layers [66]. We believe that this simplification is a reasonable price to pay for building a realistic model more closely aligned with experimentally observed structures.

There are two possible distribution patterns for functional groups on the basal plane of graphene: the first is a uniform distribution (*e.g.*, Bekyarova *et al.* [67]), and the second is the clustering of impurities to form island-like functionalised areas. The results of previous modelling of the step-by-step functionalisation of graphene by various species [68,69] and the experimentally observed inhomogeneities in the distribution of oxidised areas in GO (*e.g.*, Gomez-Navarro *et al.* [70]) suggest the prevalence of the second scenario. One more argument for the selection of this pattern of functionalisation is the different ratios of $sp^2/sp^3$ species observed in the XPS spectra of real o-CDs with the same or similar PL spectral bands (see Table I). In the case of uniform distribution, different $sp^2/sp^3$ ratios correspond to different bandgap values and, therefore, to different energies of emitted photons — in disagreement with experimental results. On the other hand, multiple theoretical works (see, *e.g.*, Xiang *et al.* [71]) suggest that nanoribbon-like $sp^2$ regions within the functionalised matrix can result in a bandgap. The values of the bandgaps of $sp^2$-nanoribbons are smaller than the values of the bandgap for functionalised areas. The optical properties of the system are predominantly defined by the highest position of the valence band (or HOMO in molecules) and the lowest position of the conduction band (LUMO in molecules). Hence, we can consider that the electronic structure of non-functionalised areas defines the optical properties of the whole system (as shown in Scheme 1a). Therefore, in further discussions we will discuss the atomic and electronic structure of the carbon hybridisation regime with smaller values of the bandgap, which defines the colour of PL. It should be noted that in the case of a junction of the conductive (gapless) *sp²* hybridised and semiconductive *sp³* hybridised regions, photogenerated charge carriers can easily recombine in the conducting region without emission of a photon (Scheme 1b). Note that, since vacancies and dislocations (*e.g.,* Stone–Wales defects) are more chemically active than perfect *sp2*-hybridised areas [68], all defective sites will be oxidised first and therefore be found preferentially in the *sp3*-hybridised part of the system.

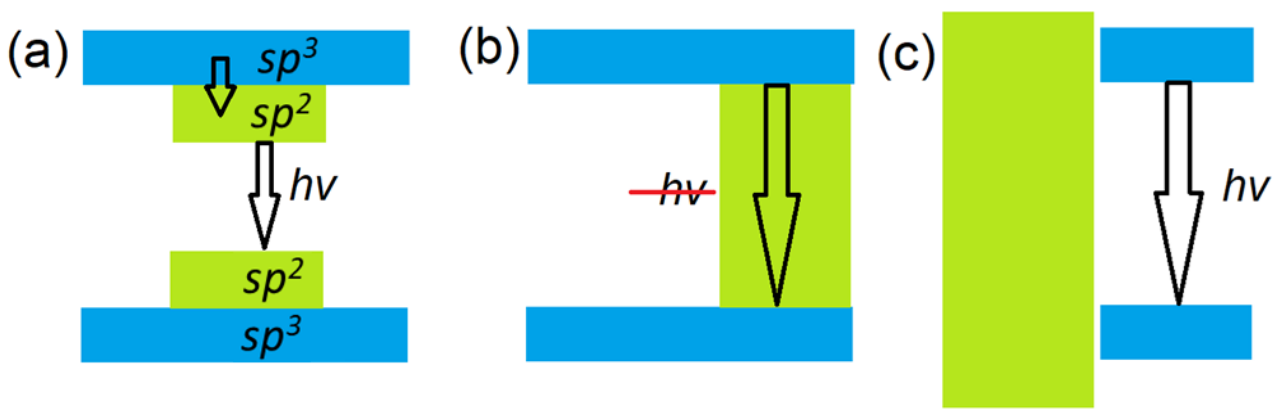

**Scheme 1.** Band diagrams and possible pathways of electron-hole recombination for the cases of: a *sp²* hybridised area incorporated in the *sp³* hybridised matrix (a) with a bandgap and (b) without a bandgap; and (c) a separate *sp³* hybridised region at the edge of a gapless *sp²* hybridised region. Examples of realistic structures corresponding with each scheme are discussed in text.

**Figure 1.** Optimised atomic structure of a *sp²*-island of 24 carbon atoms inside (a) large, (b) medium and (c) small supercells of graphane, and (d) a small supercell of graphene oxide. Carbon atoms are shown in grey, hydrogen in cyan, and oxygen in red. The same colour scheme is used in all further figures herein. (e) the densities of states of carbon atoms in the *sp²*-islands in (a-d). The Fermi energy in (e) and in similar figures throughout this work was set as zero.

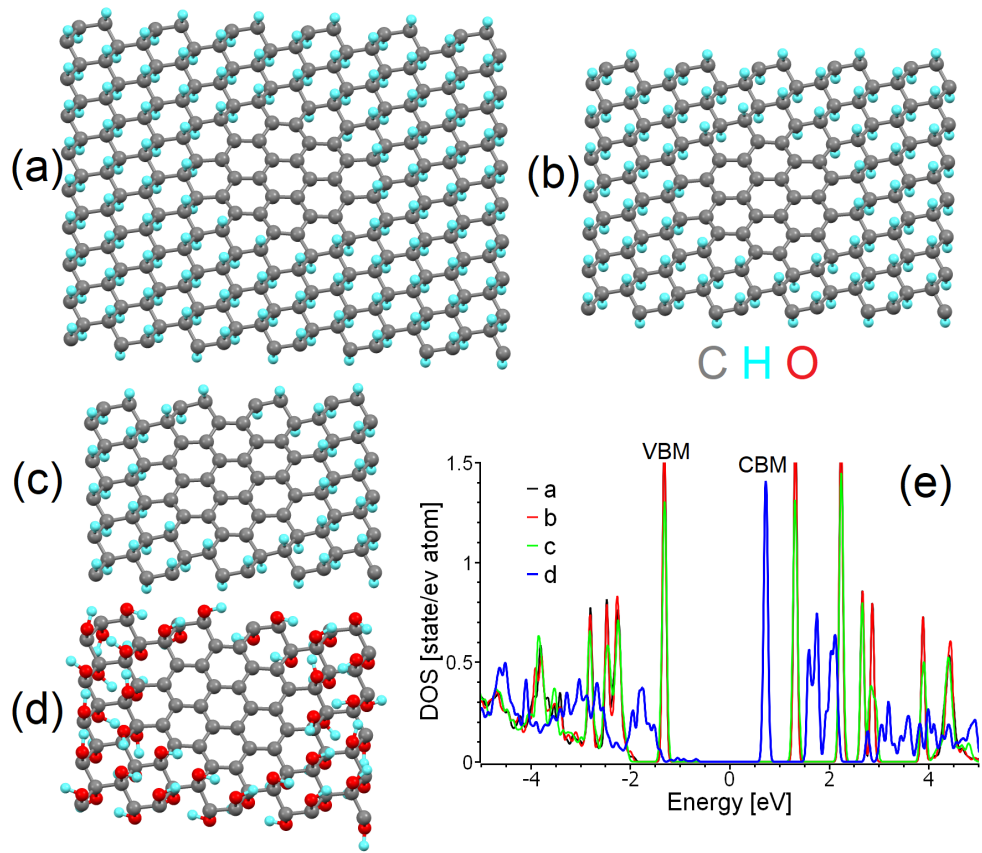

Since the computational time is proportional to the cube of the number of considered orbitals, we further simplify our model. First, we only consider the functionalisation of graphene by hydroxyl groups (Fig. 1d) and then substitute the hydroxyl groups with hydrogen atoms (Fig. 1c). This simplification provides a significant decrease in the computational cost for a supercell of the same size and with the same number of functionalised carbon atoms (carbons with $sp^3$-hybridisation). The downsides of this shortcut and methods to account for and correct the introduced inaccuracies in the simplified model are discussed further in section 4.3.

The next step is to select the correct patterns for the functionalised ($sp^3$) areas. In this case, there are again two possible scenarios. In the first, functionalised regions of limited size are lo-calised in the centre of the supercell. In the second, it is instead the non-functionalised areas (hereinafter $sp^2$-islands) that are localised there. Since functionalisation of the edges of graphene sheets is more energetically favourable than that for internal regions (holding true for various sizes and shapes of graphene sheets) [72-75], we implement the second scenario into our model. In this case, gradual oxidation of graphene from the edges towards the central area or segregation of the oxygen to areas on the edges by migration [76,77] will lead to the formation of non-functionalised $sp^2$-islands in internal regions of the basal plane (see Fig. 1a-d). The final step in building the model is to assign the size of supercell. Different sizes of the supercells correspond to different distances between nearest-neighbours $sp^2$-islands under the periodic boundary conditions imposed. To assess the effect of the distance between $sp^2$-islands on the predicted electronic structure, we constructed three different rectangular supercells of partially hydrogenated graphene (graphane) of different sizes (216, 140 and 80 carbon atoms). Identical islands of 24 $sp^2$-hybridised carbon atoms (with exactly the same shape and size) were positioned in the centre of the supercells (Fig. 1a-c). These three periodic supercells correspond with 0.92, 0.7 and 0.5 nm distances between centres of the $sp^2$-islands. The results of the calculations demonstrate that the size of the supercell (*i.e.*, the distance between $sp^2$-islands in the graphane matrix) does not play a significant role in the electronic structure of the non-functionalised areas (see Fig. 1e). Thus, the choice of the supercell can be determined by a balance between the computational cost and the desire to simulate larger $sp^2$-islands. For these reasons, we will further use the supercell shown in Fig. 1a.

**Figure 2.** (a-k) Optimised atomic structures of polycyclic aromatic hydrocarbons (PAHs) with morphologies of the carbon structures similar to $sp^2$-islands in graphane, and (m-l) graphane supercells with armchair $sp^2$-nanoribbons of different widths. The numbers under the pictures give the values of the calculated bandgaps in the PAHs or nanoribbons (first number) and in the $sp^2$-

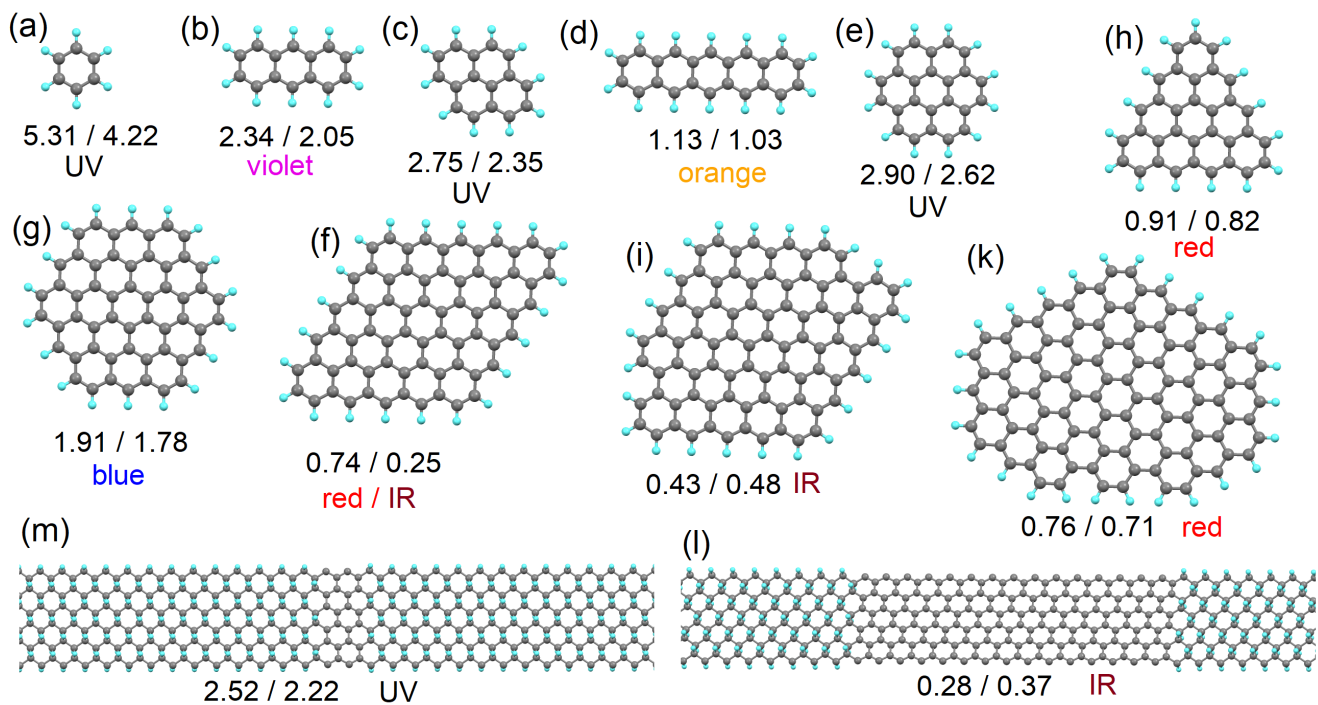

islands or *sp²*-nanoribbons with the same morphology (second number). The expected PL colour (see Table II) of each structure is also indicated.

### 4.2. *sp²*-islands vs polycyclic aromatic hydrocarbons

From the *sp²*-islands model proposed, we then look to calculate the optical properties (energy of emitted photoelectrons) at various ratios of *sp²*/*sp³* carbon and for various morphologies of the *sp²*-islands. Additionally, because the morphology of *sp²*-islands (see Fig. 1a-d) is similar to the morphology of the carbon part of polycyclic aromatic hydrocarbons (PAHs), and since PAHs are used to model the optical properties of some CQDs, we also calculate atomic and electronic structures for PAHs with carbon bodies of the same morphology as for the modelled *sp²*-islands. Full optimisation of the atomic positions was also carried out for all considered PAHs. Since PAHs provide a clear and simple picture of the of *sp²*-island morphologies (*c.f.* Fig. 2e vs

Fig. 1a), we will illustrate the morphologies of the $sp^2$-islands considered through reciprocal PAHs (Fig. 2a-k). In addition to isolated $sp^2$-islands, we also simulated several $sp^2$-nanoribbons (Fig. 2m-l) and compared the calculated values of the bandgap with the values calculated for graphene nanoribbons of the same morphology. The HOMO/LUMO gaps in the PAHs and the bandgaps of $sp^2$-islands with the same morphologies are reported in Fig. 2. The value of the bandgap of $sp^2$-islands and $sp^2$-nanoribbons is typically smaller than that calculated for reciprocal PAHs and nanoribbons. Thus, the use of PAHs as a model for o-CDs with layered structures can be performed only by introducing further corrections. The nature of the difference between the values of the bandgaps in $sp^2$-islands and PAHs and a simple approach towards correcting PAH-based models is discussed in the next section. Note that $sp^2$-islands located at the periphery of graphene are an intermediate case between $sp^2$-islands at the centre of infinite graphene sheets and PAHs. Therefore, the values of the bandgap $sp^2$-islands on the edges should also be intermediate between values calculated for PAHs and infinite graphene sheets.

The second result from the simulations is the rapid decrease observed in the values of the bandgap with increasing $sp^2$-island and PAH sizes (even accounting for the corrections discussed in section 3 and summarised in Table II). This result is similar to that reported in the literature for the modeled red shift in $sp^2$-carbon subdomains in graphene oxide [78]. Existing experimental results (see Table I) demonstrate that real nitrogen-free o-CDs usually demonstrate PL in the blue part of the spectrum, which corresponds with bandgaps in the range 1.7–1.9 eV (see Table II). These values are seen in the modelling of the smallest (below 60 carbon atoms) $sp^2$-islands or PAHs. On the other hand, the measured $sp^2/sp^3$ ratios for blue-emitting o-CDs of a few nanometres in size is from 3:1 to 20:1. Thus, the number of carbon atoms in non-oxidised areas ($sp^2$-islands) exceeds a hundred, and our proposed model only successfully predicts the properties of the green and red emitting o-CDs with significant oxidised areas (such as systems 4, 5 and 11 from Table I).

The results of the calculations presented in Fig. 1a-k, also demonstrate another significant drawback of the description of CQDs as PAHs or $sp^2$-islands in a $sp^3$-matrix: violation of hexagonal symmetry in both PAHs and $sp^2$-islands leads to a significant decrease in the values of the bandgaps. Thus, the experimentally observed blue emission corresponds with high-symmetry $sp^2$-islands and PAHs. Sometimes, the interplay between the symmetries of the $sp^2$-island and the supercell can provide an additional contribution to the discrepancy in the values of the bandgap between PAHs and $sp^2$-islands (see, *e.g.*, Fig. 2f). However, a top-down or bottom-up approach leads to the formation of PAHs or $sp^2$-islands of various shapes and sizes. The likelihood of the formation of high-symmetry blue-emitting structures should be very low. In addition, in the case of the formation of some mix of PAHs with different values of their HOMO/LUMO gaps, the scenario illustrated in Scheme 1a will be implemented. Thus, either $sp^2$-islands or PAHs are a suitable model only for o-CDs with emission from green to red.

### 4.3. Effect of Distortion on the Bandgaps of $sp^2$-islands and Nanographenes

We have thus far discussed only $sp^2$-islands on perfectly flat graphene sheets. However, graphene sheets usually deviate from a flat geometry [79]. Peculiarities of nanoparticle morphologies such as impurities, solvent molecules, or intercalants can also contribute to the deformation of graphenic membranes geometries. Therefore, one should also take into account the effect of out-of-plane corrugation on the electronic structure of $sp^2$-islands. To do so, we used a further approach where the in-plane lattice parameters of the supercell were reduced (in this work by 10%) to give an initial out-of-plane deviation of some atoms in the centre of the supercell (in this work all atoms of the $sp^2$-island were shifted up by 0.01 nm).

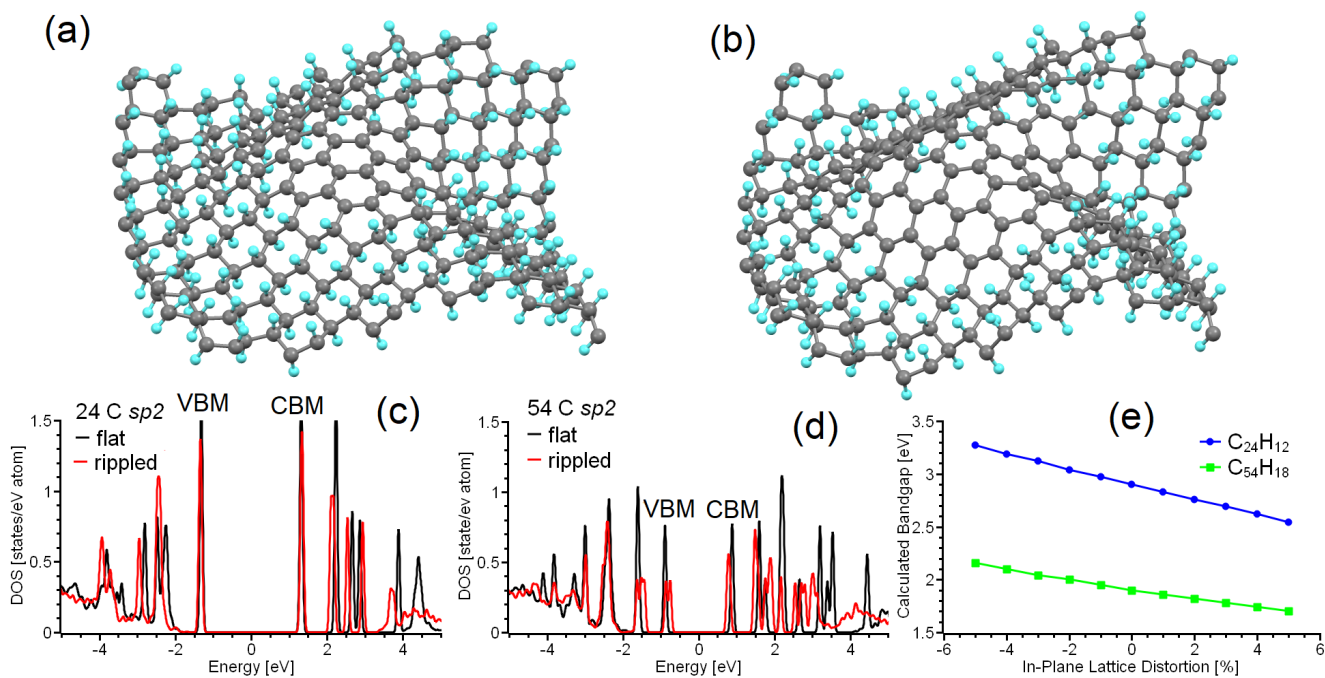

**Figure 3.** Optimised atomic structures of corrugated graphene with $sp^2$-islands of (a) 24 and (b) 54 carbon atoms, and densities of states for $sp^2$-islands of (c) 24 and (d) 54 in flat and rippled graphane. (e) The changes of the bandgap of $C_{24}H_{12}$ (as given in Fig. 2e) and $C_{54}H_{18}$ (as given in Fig. 2g) as a function of the change of in-plane lattice parameter. The Fermi energy in (c) and (d) was set as zero.

We further performed optimisation of the atomic position with fixed lattice parameters. To study the effect of the corrugation on the electronic structure we choose the specific cases of the graphene supercells with coronene-like $sp^2$-islands of 24 and 54 carbon atoms and compare electronic structure of $sp^2$-islands before (Fig. 1a) and after (Fig. 3a,b) corrugation. The results of the calculations demonstrate the presence of a redshift in the electronic structure of the $sp^2$-island with 24 carbon atoms (Fig. 3c) and a larger redshift for the $sp^2$-island with 54 atoms in the corrugated matrix. Thus, the corrugation of graphenic sheets leads to a redshift in the spectrum of the

$sp^2$-matrix. Note that substitution of the hydrogen by hydroxyl groups or fluorine leads to larger corrugated distortions of the graphene surface [42,58]. Hence, the value of the bandgap of $sp^2$-islands in a GO-like matrix is smaller that in an otherwise equivalent graphane environment (see Fig. 1d).

Based on the simulation of the effect of lattice distortion on the electronic structure of the $sp^2$-islands, we also considered the effect of in-plane strain on the electronic structure of the PAHs. For this purpose, we performed optimisation of the atomic structures of $C_{24}H_{12}$ (Fig. 2e) and $C_{54}H_{18}$ (Fig. 2g) and then varied the in-plane lattice parameter from 95% to 105% of the initial value without further optimisation of the atomic positions. The results demonstrated that in-plane compression of PAHs leads to a blue shift and expansion to a redshift. This result can be used to simplify the modelling of $sp^2$-islands in nanographenes. First, several initial PAHs with values of their bandgaps closer to experimental equivalents should be chosen. Second, based on experimental characterisation of the chemical composition of functionalised graphene, the magnitude of the changes to the lattice structure after functionalisation can be calculated. The obtained values of in-plane lattice distortions should be used for planar compression or expansion of the considered PAHs in the model. These deformed PAHs can be used in further calculations (similar to those performed by Kundelev *et al.* [23]) of the optical properties of o-CDs. The described approach significantly decreases the computational costs by reducing supercells of functionalised graphene with $sp^2$-islands to PAHs. This improves the availability and access towards time-dependent DFT-based calculations based on realistic chemical compositions of the simulated o-CDs.

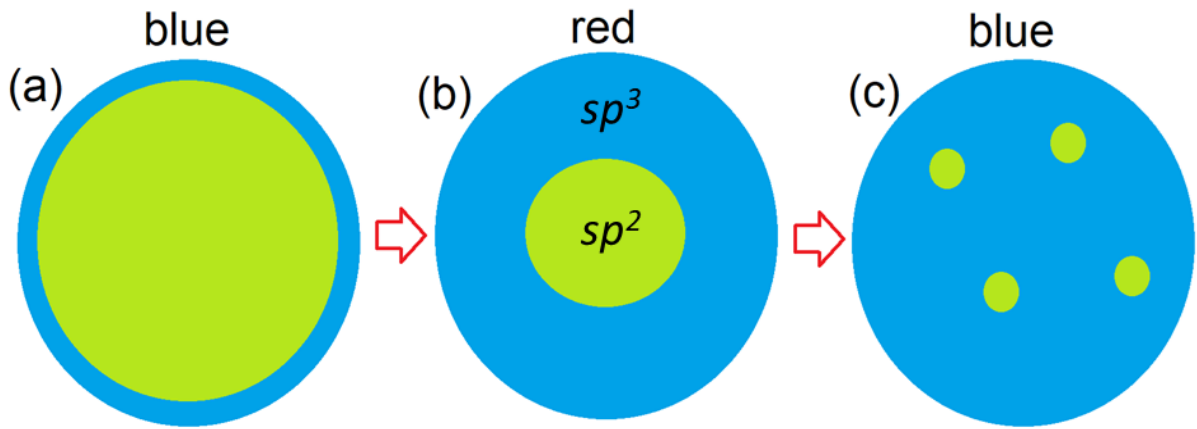

**Scheme 2.** A sketch of a nanographene domain with (a) oxidised ($sp^3$) carbon on the edges, (b) a single large $sp^2$-island in the centre, and (c) several small $sp^2$-islands in the centre. The most likely PL colours of emitting light are shown above the panels. Blue areas represent oxidised parts and green areas denote non-oxidised graphene.

## 5. Oxidised Edges of Nanographenes

We have so far explored and discussed the limitations of $sp^2$-islands and PAHs as models of realistic nitrogen-free o-CDs. However, all the previously described models are suitable only for graphenic sheets of sub-nanometre sizes or for o-CDs with a significant level of oxidation ($sp^2/sp^3 < 1$). Note that only a two of the samples shown in Table I (*i.e.*, 6 and 7) satisfy these conditions. Thus, some different model system is required for o-CDs with low amounts of oxygen in the basal plane. Taking into account the functionalisation of graphene sheets from the edges to the centre as discussed in section 4.1 and also in [61,65], we considered the influence of the oxidation of areas at the edge of the graphenic sheet upon the electronic structure. For this purpose, we selected a test case nanographene of larger size (104 carbon atoms) than that previously considered (Fig. 1k). Since the values of the energy gaps in graphene oxide functionalised with only hydroxyl groups and with a mix of epoxy and hydroxyl groups are sufficiently similar [52], we consider only the latter in our further modeling. We then defined a functionalised (with hydroxyl and carbonyl groups) region of twelve carbon atoms at the edge (see Fig. 4a). The distribution of the functional groups was chosen based on the principles previously described for graphene oxide [52] and for the oxidised edges of graphene nanoribbons [68,76]. This initial structure corresponds to an 8:1 ratio of $sp^2$ to $sp^3$ carbon atoms in the system. We then gradually increased the size of the $sp^3$-island by the iterative addition of four hydroxyl groups (two from one side of the plane and two from the other) until finally reaching a $sp^2$:$sp^3$ ratio of 3:1 (Fig. 4d). Note that the range of $sp^2$:$sp^3$ ratios spanned by these simulations is similar to that found experimentally in real CQDs (see Table I).

The optical properties of these oxidised regions are critically defined by the two distinct peaks below and above Fermi level, with a peak-to-peak separation of around 2 eV (see the right column of Fig. 4), corresponding with PL in the blue part of the spectrum (see Table II). The minor peaks near the Fermi level (and between the two main peaks of interest) result from the interaction of the functionalised area with nano-graphene of size smaller than that observed in experiments (see Table I). Note that, in contrast to $sp^2$-islands and PAHs where morphology significantly influences the value of the bandgap (see Fig. 2), in the case of $sp^3$-islands (oxidised areas) the value of the bandgap is defined only by the chemical composition of the functional groups and almost does not depend on the shape or size of these islands. The results presented here explain how the o-CDs of different sizes and with different $sp^2$:$sp^3$ ratios summarised in Table I (*i.e.*, 1, 5, 6, 10, 11, 13, and 15) demonstrate PL at the same or almost the same wavelength. The insignificance of the shape and size of the oxidised area on the value of bandgap is due to the formation of a diamond-like 2D structure. Changes in the shape and size, or substitution of hydroxyl groups for epoxy groups leads to a mild in-plane strain, which in turn causes only limited changes to the values of the bandgap [52]. Therefore, the smallest fully oxidised nanographene can be used as a model of oxidised regions on the edges of nanographenes for further modelling of the optical properties of o-CDs.

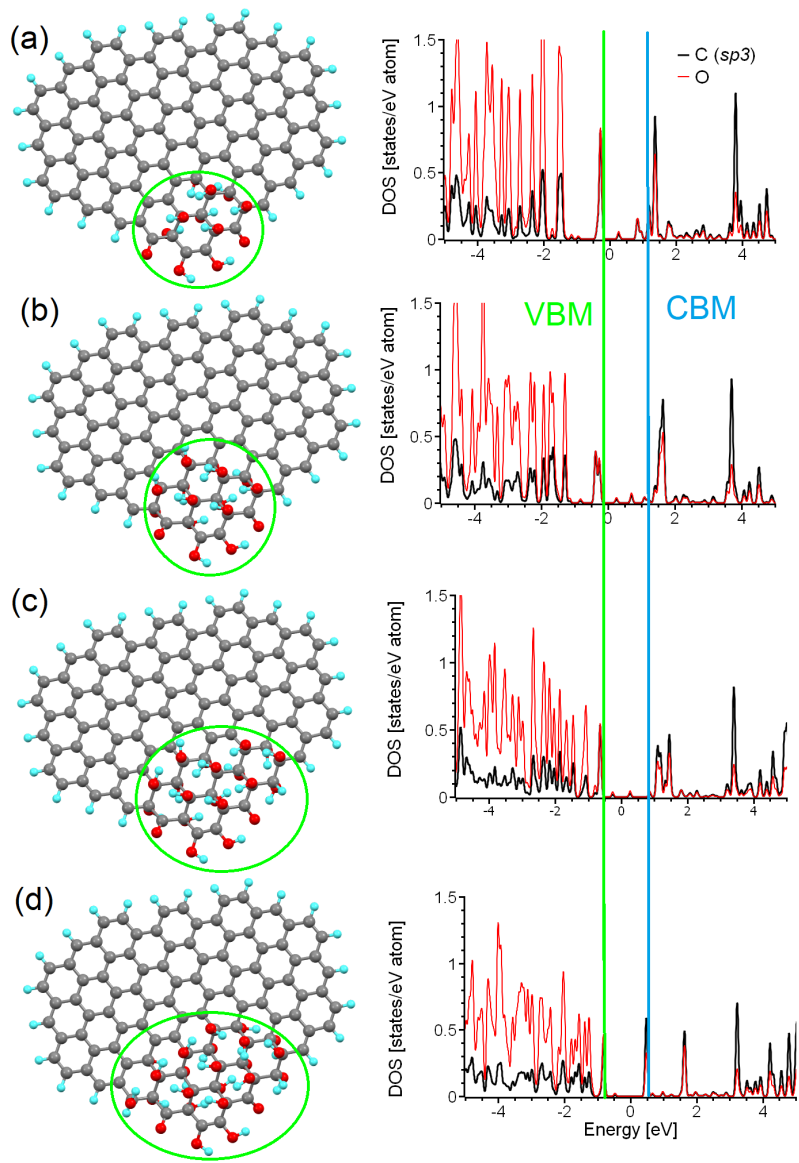

**Figure 4.** (Left) Optimised atomic structure of the same nanographene with different sizes of the oxidised ($sp^3$) island at the edge (shown by green circles), and (Right) corresponding partial densities of states of carbon and oxygen atoms belonging to these $sp^3$-islands. Approximate $sp^2$:$sp^3$ ratios of (a) 8:1, (b) 6:1, (c) 4:1, and (d) 3:1 are considered. The Fermi energies for all four cases was set as zero.

**6. Conclusions and Outlook**

Based on the results of modelling experimentally characterised, graphene-based, carbon quantum dots, we can propose further schemes for the atomic structures of these systems. For low oxygen content ($sp^2$:$sp^3$ >> 1), oxidised regions are formed at the edges of the graphenic sheets (Scheme 2a). To simulate the optical properties of these systems, nanographenes with oxidised edges (similar to those shown in Fig. 4a) can be used. An increase of the oxygen content leads to the formation of smaller, but still large (around 1 nm) graphenic $sp^2$-islands in a $sp^3$-matrix (Scheme 2b and Fig. 1a-d). Changes in the bandgap values along this route can be accurately described by the simple semi-empirical formula proposed by Wei *et al.* [39]. A further increase of oxygen content in the basal plane of the graphene sheets leads to further diminishing in the size of graphenic $sp^2$-islands (Scheme 2c) with a corresponding blue shift in the PL. An accurate theoretical description and prediction of optical absorption and fluorescent emission properties of these $sp^2$-islands can be obtained from model systems of strained polycyclic aromatic hydrocarbons. Thus, based on the comparison of known experimental data and simulations of nitrogen-free CQDs, we can propose the oxidation and reduction of synthesised nanoparticles as the simplest methods for fine-tuning the optical properties of these systems. The ratio of $sp^2$ to $sp^3$ measured in the synthesised CQDs provides enough information to determine a starting point, and the models presented herein will further guide treatment (oxidation or reduction, where the carbon atoms in the graphene nanosheet undergo rehybridisation between $sp^2$ and $sp^3$ states) towards obtaining the desired PL properties. Importantly, we have shown that there is no need to involve ideas about elementary emitters in the form of molecules of polycyclic aromatic compounds on the surface/periphery of the particles to explain the light emission by CQD particles in the 400–600 nm range (even in the red part of the spectrum, as for compounds 8 and 9 from Tab. I). The basic CQD particles themselves can be up to 8-10 nm (*i.e.,* not necessarily small (up to 2 nm)). The developed models are in good agreement with the experimental data on the properties of CQDs available from the literature.

**Acknowledgements**

D.W.B. acknowledges support from the Russian Science Foundation (Project No 21-12-00392). V.Yu.O. acknowledges the support from Ioffe Institute (Project 0040-2014-0013).

**References**

1) T. Enoki, K. Takai, V. Osipov, M. Baidakova, A.Vul', Nanographene and Nanodiamond; New Members in the Nanocarbon Family. *Chem. Asian J.* 2009, **4**, 796-804. https://doi.org/10.1002/asia.200800485

2) X.Y. Xu, R. Ray, Y.L. Gu, H.J. Ploehn, L. Gearheart, K. Raker, W.A. Scrivens, Electrophoretic analysis and purification of fluorescent single-walled carbon nanotube fragments. *J. Am. Chem. Soc.* 2004, **126**, 12736–12737. https://doi.org/10.1021/ja040082h

3) Y.-P. Sun, B. Zhou, Y. Lin, W. Wang, K. A. S. Fernando, P. Pathak, M. J. Meziani, B. A. Harruff, X. Wang, H. Wang, P. G. Luo, H. Yang, M. E. Kose, B. Chen, L. M. Veca, S.-Y. Xie, Quantum-sized carbon dots for bright and colorful photoluminescence. *J. Am. Chem. Soc.* 2006, **128**, 7756. https://doi.org/10.1021/ja062677d

4) D. Tang, J. Liu, X. Wu, R. Liu, X. Han, Y. Han, X. Han, Y. Liu, Z. Kang, Carbon quantum dot/nife layered double-hydroxide composite as a highly efficient electrocatalyst for water oxidation. *ACS Appl. Mater. Interfaces* 2014, **6**, 7918–7925. https://doi.org/10.1021/am501256x

5) M.J. Molaei, A review on nanostructured carbon quantum dots and their applications in biotechnology, sensors, and chemiluminescence. *Talanta* 2019, **196**, 456. https://doi.org/10.1016/j.talanta.2018.12.042

6) M.J. Molaei, Carbon quantum dots and their biomedical and therapeutic applications: a review. *RSC Adv.* 2019, **9**, 6460. https://doi.org/10.1039/C8RA08088G

7/6) X. Wang, Y. Feng, P. Dong, J. Huang, A Mini Review on Carbon Quantum Dots: Preparation, Properties, and Electrocatalytic Application. *Front. Chem.* 2019, **7**, 671. https://doi.org/10.3389/fchem.2019.00671

8) M.J. Molaei, Principles, mechanisms, and application of carbon quantum dots in sensors: a review. *Anal. Meth.* 2020, **12**, 1266. https://doi.org/10.1039/C9AY02696G

9) M.J. Molaei, The optical properties and solar energy conversion applications of carbon quantum dots: A review. *Solar Ener.* 2020, **196**, 549. https://doi.org/10.1016/j.solener.2019.12.036

10) A.S. Rasal, S. Yadav, A. Yadav, A.A. Kashale, S.T. Manjunatha, A. Altaee, J.-Y. Chang, Carbon Quantum Dots for Energy Applications: A Review. *ACS Appl. Nano Mater.* 2021, **4**, 6515. https://doi.org/10.1021/acsanm.1c01372

11) S. Sharma, V. Dutta, P. Singh, P. Raizada, A. Rahmani-Sani, A. Hosseini-Bandegharaei, V. K. Thakur, Carbon quantum dot supported semiconductor photocatalysts for efficient degradation of organic pollutants in water: A review. *J. Cleaner Product.* 2019, **228**, 755. https://doi.org/10.1016/j.jclepro.2019.04.292

12) K.J. Mintz, Y. Zhou, R.M. Leblanc, Recent development of carbon quantum dots regarding their optical properties, photoluminescence mechanism, and core structure. *Nanoscale* 2019, **11**, 4634. https://doi.org/10.1039/C8NR10059D

13) R. Das, R. Bandyopadhyay, P. Pramanik, Carbon quantum dots from natural resource: A review. *Mater. Today Chem.* 2018, **8**, 96. https://doi.org/10.1016/j.mtchem.2018.03.003

14) H. Fan, M. Zhang, B. Bhandari, C. Yang, Food waste as a carbon source in carbon quantum dots technology and their applications in food safety detection. *Trends in Food Sci. & Technol.* 92020, **5**, 86. https://doi.org/10.1016/j.tifs.2019.11.008

15) X. Wang, Y. Feng, P. Dong, J. Huang, A Mini Review on Carbon Quantum Dots: Preparation, Properties, and Electrocatalytic Application. *Front. Chem.* 2019, **7**, 671. https://doi.org/10.3389/fchem.2019.00671

16) J. Ren, L. Malfatti, P. Innocenzi, Citric Acid Derived Carbon Dots, the Challenge of Understanding the Synthesis-Structure Relationship. *C* 2021, **7**, 2. https://doi.org/10.3390/c7010002

17) J. Lim et al., *Nat Commun.* 2016, **7**, 10364. https://doi.org/10.1038/ncomms10364

18) N.V. Tepliakov, E.V. Kundelev, P.D. Khavlyuk, Y. Xiong, M.Yu. Leonov, W. Zhu, A.V. Baranov, A.V. Fedorov, A.L. Rogach, I.D. Rukhlenko, sp2–sp3-Hybridized Atomic Domains Determine Optical Features of Carbon Dots. *ACS Nano* 2019, **13**, 10737–10744. https://doi.org/10.1021/acsnano.9b05444

19) T. Li, Z. Li, T. Huang, L. Tian, Carbon quantum dot-based sensors for food safety. *Sens. Act. A: Phys.* 2021, **331**, 113003. https://doi.org/10.1016/j.sna.2021.113003

20/16) S. Zhu, Y. Song, X. Zhao, J. Shao, J. Zhang, B. Yang, B. (2015). The photoluminescence mechanism in carbon dots (graphene quantum dots, carbon nanodots, and polymer dots): current state and future perspective. *Nano Res.* 2015, **8**, 355–381. https://doi.org/10.1007/s12274-014-0644-3

21) B. Zhi, et. al., Multicolor polymeric carbon dots: synthesis, separation and polyamide-supported molecular fluorescence. *Chem. Sci.*, 2021, **12**, 2441-2455. https://doi.org/10.1039/D0SC05743F

22) S. Li, L. Li, H. Tu, H. Zhang, D.S. Silvester, C.E. Banks, G. Zou, H. Hou, X. Jia, The development of carbon dots: From the perspective of materials chemistry. *Mater. Today* 2021, **51**, 188-207. https://doi.org/10.1016/j.mattod.2021.07.028

23) B.B. Chen, M.L. Liu, C.M. Li, C.Z. Huang, Fluorescent carbon dots functionalization. *Adv. Coll. Interf. Sci.* 2019, **270**, 165-190. https://doi.org/10.1016/j.cis.2019.06.008

24) K.J. Mintz, et al., A deep investigation into the structure of carbon dots. *Carbon* 2021, **173**, 433-447. https://doi.org/10.1016/j.carbon.2020.11.017

25) V. R. Galakhov, A. Buling, M. Neumann, N. Ovechkina, A. Shkvarin, A. Semenova, M. Uimin, A. E. Yermakov, E. Z. Kurmaev, O. Vilkov, D. W. Boukhvalov, Carbon States in Carbon-Encapsulated Nickel Nanoparticles Studied by Means of X-Ray Absorption, Emission, and Photoelectron Spectroscopies. *J. Phys. Chem. C* 2011, **115**, 24615. https://doi.org/10.1021/jp2085846

26) R. Das, N. Dhar, A. Bandyopadhyay, D. Jana, Size dependent magnetic and optical properties in diamond shaped graphene quantum dots: A DFT study. *J. Phys. Chem. Solids* 2016, **99**, 34. https://doi.org/10.1016/j.jpcs.2016.08.004

27) E.V. Kundelev, N.V. Tepliakov, M.Yu. Leonov, V. G. Maslov, A.V. Baranov, A.V. Fedorov, I.D. Rukhlenko, A.L. Rogach Toward Bright Red-Emissive Carbon Dots through Controlling Interaction among Surface Emission Centers. *J. Phys. Chem. Lett.* 2020, **11**, 8121. https://dx.doi.org/10.1021/acs.jpclett.0c02373

28) EV. Kundelev, E.D. Strievich, N.V. Tepliakov, A.D. Murkina, A.Yu. Dubavik, E.V. Ushakova, A.V. Baranov, A.V. Fedorov, I.D. Rukhlenko, A.L. Rogach, Structure–Optical Property Relationship of Carbon Dots with Molecular-like Blue-Emitting Centers, *J. Phys. Chem. C* 2022, **126**, 18170–18176. https://doi.org/10.1021/acs.jpcc.2c05926

29) J. Xu, Q. Liang, Z.Li,V.Yu. Osipov, Y. Lin, B. Ge,Q. Xu, J.Zhu, H. Bi, Rational Synthesis of Solid-State Ultraviolet B Emitting Carbon Dots via Acetic Acid-Promoted Fractions of $sp^3$ Bonding Strategy. *Adv. Mater.* 2022, **34**, 20222200011. https://doi.org/10.1002/adma.202200011

30) A. Dager, T. Uchida, T. Maekawa, M. Tachibana, Synthesis and characterization of Monodisperse Carbon Quantum Dots from Fennel Seeds: Photoluminescence analysis using Machine Learning. *Sci. Rep.* 2019, **9**, 14004. https://doi.org/10.1038/s41598-019-50397-5

31) A. Dager, A. Baliyan, S. Kurosu, T. Maekawa, M. Tachibana, Ultrafast synthesis of carbon quantum dots from fenugreek seeds using microwave plasma enhanced decomposition: application of C-QDs to grow fluorescent protein crystals. *Sci. Rep.* 2020, **10**, 12333. https://doi.org/10.1038/s41598-020-69264-9

32) S.S. Jones, P. Sahatiya, S. Badhulik, One step, high yield synthesis of amphiphilic carbon quantum dots derived from chia seeds: a solvatochromic study. *New J. Chem.* 2017, **41**, 13130. https://doi.org/10.1039/C7NJ03513F

33) M.Y. Pudza, Z.Z. Abidin, S.A. Rashid, F.M. Yasin, A.S.M Noor, M.A. Issa, Eco-Friendly Sustainable Fluorescent Carbon Dots for the Adsorption of Heavy Metal Ions in Aqueous Environment. Nanomaterials 10 (2020) 315. https://doi.org/10.3390/nano10020315

34) M. He, J. Zhang, H. Wang, Y. Kong, Y. Xiao, W. Xu, Material and Optical Properties of Fluorescent Carbon Quantum Dots Fabricated from Lemon Juice via Hydrothermal Reaction. *Nanoscale Res. Lett.* 2018, **13**, 175. https://doi.org/10.1038/s41598-020-69264-9

35) G.L. Hong, H.L. Zhao, H.H. Deng. H.J. Yang, H.P. Peng, Y.H. Liu, W. Chen, W. Fabrication of ultra-small monolayer graphene quantum dots by pyrolysis of trisodium citrate for fluorescent cell imaging. *Int. J. Nanomed.* 2018, **13**, 4807. https://doi.org/10.2147/IJN.S168570

36) X. Ma, S. Li, V. Hessel, L. Lin, S. Meskers, F. Gallucci, Synthesis of luminescent carbon quantum dots by microplasma process. Chem. Eng. Processing - *Process Intensification.* 2019, **140**, 29. https://doi.org/10.1016/j.cep.2019.04.017

37) Z. Wang, F. Yuan, X. Li, Y. Li, H. Zhong, L. Fan, S. Yang, S. 53% Efficient Red Emissive Carbon Quantum Dots for High Color Rendering and Stable Warm White-Light-Emitting Diodes. *Adv. Mater.* 2017, **29**, 1702910. https://doi.org/10.1002/adma.201702910

38) S. Sarkar, D. Gandla, Y. Venkatesh, P.R. Bangal, S. Ghosh, Y. Yang, S. Misra, Graphene quantum dots from graphite by liquid exfoliation showing excitation-independent emission, fluorescence upconversion and delayed fluorescence. *Phys. Chem. Chem. Phys.* 2016, **18**, 21278. https://doi.org/10.1039/C6CP01528J

39) J. Peng et. al., Graphene Quantum Dots Derived from Carbon Fibers. *Nano Lett.* 2012, **12**, 844. https://doi.org/10.1021/nl2038979

40) Y. Zhang, H. Gao, J. Niu, B. Liu, Facile synthesis and photoluminescence of graphene oxide quantum dots and their reduction products. *New J. Chem.* 2014, **38**, 4970. https://doi.org/10.1039/C4NJ01187B

41) S. Jeong, R.L. Pinals, B. Dharmadhikari, H. Song, A. Kalluri, D. Debnath, Q. Wu, M.-H. Ham, P. Patra, M.P. Landry, Graphene Quantum Dot Oxidation Governs Noncovalent Biopolymer Adsorption. *Sci. Rep.* 2020, **10**, 7074. https://doi.org/10.1038/s41598-020-63769-z

42) S. Tachi, H. Morita, M. Takahashi, Y. Okabayashi, T. Hosokai, T. Sugai, S. Kuwahara, Quantum Yield Enhancement in Graphene Quantum Dots via Esterification with Benzyl Alcohol. *Sci. Rep.* 2019, **9**, 14115. https://doi.org/10.1038/s41598-019-50666-3

43) F. Yuan, L. Ding, Y. Li, X. Li, L. Fan, S. Zhou, D. Fang, Multicolor fluorescent graphene quantum dots colorimetrically responsive to all-pH and a wide temperature range. *Nanoscale* 2015, **7**, 11727. https://doi.org/10.1039/C5NR02007G

44) W. Zhang, Y. Liu, X. Meng, T. Ding, Y. Xu, H. Xu, Y. Ren, B. Liu, J. Huang, J.; Yang, X. Fang, Graphenol defects induced blue emission enhancement in chemically reduced graphene quantum dots. *Phys. Chem. Chem. Phys.* 2015, **17**, 22361. https://doi.org/10.1039/C5CP03434E

45) K. Wei, F. Liao, H. Huang, M. Shao, H. Lin, Y. Liu, Z. Kang, Simple Semiempirical Method for the Location Determination of HOMO and LUMO of Carbon Dots. *J. Phys. Chem. C* 2021, **125**, 7451. https://doi.org/10.1021/acs.jpcc.1c00812

46) W. Shi, Q. Han, J. Wu, C. Ji, Y. Zhou, S. Li, L. Gao, R.M. Leblanc, Z. Peng, Open AccessArticleSynthesis Mechanisms, Structural Models, and Photothermal Therapy Applications of Top-Down Carbon Dots from Carbon Powder, Graphite, Graphene, and Carbon Nanotubes. *Int. J. Mol. Sci.*, 2022, **23**, 1456. https://doi.org/10.3390/ijms23031456

47) K. Krishnamoorthy, M. Veerapandi, K. Yun, S.-J. Kim, The chemical and structural analysis of graphene oxide with different degrees of oxidation. *Carbon* 2013, **53**, 38-49. https://doi.org/10.1016/j.carbon.2012.10.013

48) Y.C.G. Kwan, G.M. Ng, C.H.A. Huan, Identification of functional groups and determination of carboxyl formation temperature in graphene oxide using the XPS O 1s spectrum. *Thin Solid Films* 2015, **590**, 40-48. https://doi.org/10.1016/j.tsf.2015.07.051

49) B. Wang, Z. Wei, L. Sui, J. Yu, B. Zhang, X. Wang, S. Feng, H. Song, X. Yong, Y. Tian, B.Yang S. Lu, Electron–phonon coupling-assisted universal red luminescence of o-phenylenediamine-based carbon dots. *Light Sci Appl.* 2022, **11**, 172. https://doi.org/10.1038/s41377-022-00865-x

50) H. Wu, S. Lu, B. Yang, Carbon-Dot-Enhanced Electrocatalytic Hydrogen Evolution. *Acc. Mater. Res.* 2022, **3**, 319–330. https://pubs.acs.org/doi/abs/10.1021/accountsmr.1c00194

51) X.Yang, L. Ai, J. Yu, G.I.N. Waterhouse, L.Sui, J. Ding, B. Zhang, X. Yong, S. Lu, Photoluminescence mechanisms of red-emissive carbon dots derived from non-conjugated molecules. *Sci. Bull.* 2022, **67**, 1450-1457. https://doi.org/10.1016/j.scib.2022.06.013

52) D.W. Boukhvalov, M.I. Katsnelson, Modeling of graphite oxide. *J. Am. Chem. Soc.* 2008, **130**, 10697. https://doi.org/10.1021/ja8021686

53) I. Jung, D.A. Dikin, R.D. Piner, R.S. Ruoff, Tunable Electrical Conductivity of Individual Graphene Oxide Sheets Reduced at "Low" Temperatures. *Nano Lett.* 2008, **8**, 4283. https://doi.org/10.1021/nl8019938

54) R. Das, N. Dhar, A. Bandyopadhyay, D. Jana, Size dependent magnetic and optical properties in diamond shaped graphene quantum dots: A DFT study. *J. Phys. Chem. Solids* 2016, **99**, 34-42. http://dx.doi.org/10.1016/j.jpcs.2016.08.004

55) J. Deb, D. Paul, U. Sarkar, Density Functional Theory Investigation of Nonlinear Optical Properties of T-Graphene Quantum Dots. *J. Phys. Chem. A* 2020, **124**, 1312-1320. https://dx.doi.org/10.1021/acs.jpca.9b10241

56) M.A. Sk, A. Ananthanarayanan, L. Huang, K.H. Lim, P. Chen, Revealing the tunable photoluminescence properties of graphene quantum dots. *J. Mater. Chem. C* 2014, **2**, 6954. https://dx.doi.org/10.1039/c4tc01191k

57) M. Zhao, F. Yang, Y. Xue, D. Xiao, Y. Guo, A Time-Dependent DFT Study of the Absorption and Fluorescence Properties of Graphene Quantum Dots. *ChemPhysChem* 2014, **15**, 950 -957. https://doi.org/10.1002/cphc.201301137

58) T. Jadoon, T. Mahmood, K. Ayub, DFT study on the sensitivity of silver-graphene quantum dots for vital and harmful analytes. *J. Phys. Chem. Solids* 2021, **153**, 110028. https://doi.org/10.1016/j.jpcs.2021.110028

59) H. Abdelsalam, H. Elhaes, M.A. Ibrahim, First principles study of edge carboxylated graphene quantum dots. *Phys. B: Cond. Matter* 2018, **537**, 77-86. https://doi.org/10.1016/j.physb.2018.02.001

60) J.M. Soler, E. Artacho, J.D. Gale, A. Garsia, J. Junquera, P. Orejon, D. Sanchez-Portal. The SIESTA Method for Ab-Initio Order-N Materials Simulation. *J. Phys.: Condens. Matter.* 2002, **14**, 2745. https://doi.org/10.1088/0953-8984/14/11/302

61) J.P. Perdew, K. Burke, M. Ernzerhof. Generalized Gradient Approximation Made Simple. *Phys. Rev. Lett.* 1996, **77**, 3865. https://doi.org/10.1103/PhysRevLett.77.3865

62) O.N. Troullier, J.L. Martins. Efficient Pseudopotentials for Plane-Wave Calculations. *Phys. Rev. B* 1991, **43**, 1993. https://doi.org/10.1103/PhysRevB.43.1993

63) C.J. Cramer, Essentials of computational chemistry. Theory and Models. Wiley (2004) ISBN 0-0470-09181-9 (Chapter 8).

64) M. van Schilfgaarde, Takao Kotani, and S. Faleev, Quasiparticle Self-Consistent GW Theory. *Phys. Rev. Lett.* 2006, **96**, 226402. https://doi.org/10.1103/PhysRevLett.96.226402

65) J. Heyd, G.E. Scuseria, Efficient hybrid density functional calculations in solids: Assessment of the Heyd-Scuseria-Ernzerhof screened Coulomb hybrid functional. *J. Chem. Phys.* 2004, **121**, 1187. https://doi.org/10.1063/1.1760074

66) M. Paloncýová, M. Langer, M. Otyepka, Structural Dynamics of Carbon Dots in Water and N,N-Dimethylformamide Probed by All-Atom Molecular Dynamics Simulations. *J. Chem. Theory Comput.* 2018, **14**, 2076–2083. https://doi.org/10.1021/acs.jctc.7b01149

67) E. Bekyarova, M.E. Itkis, P. Ramesh, R.C. Haddon, Chemical approach to the realization of electronic devices in epitaxial graphene. *phys. stat. solidi-rrl* 2009, **3**, 184. https://doi.org/10.1002/pssr.200903110

68) D.W. Boukhvalov, M.I. Katsnelson, Chemical functionalization of graphene with defects. *Nano Lett.* 2008, **8**, 4373-4379. https://doi.org/10.1021/nl802234n

69) D.W. Boukhvalov, Absence of a stable atomic structure in fluorinated graphene. *Phys. Chem. Chem. Phys.* 2016, **18**, 13287. https://doi.org/10.1039/C6CP01631F

70) C. Gomez-Navarro, J.C. Meyer, R.S. Sundaram, A. Chuvilin, S. Kurasch, M. Burghard, K. Kern, U. Kaiser, Atomic Structure of Reduced Graphene Oxide. *Nano Lett.* 2010, **10**, 1144.https://doi.org/10.1021/nl9031617

71) H. Xiang, E. Kan, S.-H. Wei, M.-H. Whangbo, J. Yang, "Narrow" Graphene Nanoribbons Made Easier by Partial Hydrogenation. *Nano Lett.* 2009, **9**, 12, 4025. https://doi.org/10.1021/nl902198u

72) D.W. Boukhvalov, M.I. Katsnelson, Chemical functionalization of graphene. *J. Phys.: Cond. Matter* 2009, **21**, 344205. ihttps://doi.org/10.1088/0953-8984/21/34/344205

73) D.W. Boukhvalov, X. Feng, K. Mullen, First-principles modeling of the polycyclic aromatic hydrocarbons reduction. *J. Phys. Chem. C* 2011, **115**, 16001. https://doi.org/10.1021/jp2024928

74) E.F. Sheka, N.A. Popova, Molecular theory of graphene oxide. *Phys. Chem. Chem. Phys.* 2013, **15**, 13304-13322. https://doi.org/10.1039/C3CP00032J

75) B.S. Razbirin, N.N. Rozhkova, E.F. Sheka, D.K. Nelson, A.N. Starukhin, Fractals of graphene quantum dots in photoluminescence of shungite. *J. Exp. Theor. Phys.* 2014, **118**, 735. https://doi.org/10.1134/S1063776114050161

76) D.W. Boukhvalov, Repair of magnetism in oxidized graphene nanoribbons. *Chem. Phys. Lett.* 2011, **501**, 396. https://doi.org/10.1016/j.cplett.2010.11.023

77) D.W. Boukhvalov, Modeling of hydrogen and hydroxyl group migration on graphene. *Phys. Chem. Chem. Phys.* 2010, **12**, 15367. https://doi.org/10.1039/1463-9084/1999

78) H. Yoon, Y.H. Chang, S.H. Song, E.-S. Lee, S.H. Jin, C. Park, J. Lee, B.H. Kim, H.J. Kang, Y.-H. Kim, S. Jeon, Intrinsic Photoluminescence Emission from Subdomained Graphene Quantum Dots. *Adv. Mater.* 2016, **28**, 5255-5261. https://doi.org/10.1002/adma.201600616

79) S. Deng, V. Berry, Wrinkled, rippled and crumpled graphene: an overview of formation mechanism, electronic properties, and applications. *Mater. Today* 2016, **19**, 197. https://doi.org/10.1016/j.mattod.2015.10.002